\documentclass[aps,pra,preprint,groupedaddress,showpacs]{revtex4}
\usepackage{graphicx}
\usepackage{dcolumn}
\usepackage{bm}
\begin{document}
\title{A multi-photon magneto-optical trap}
\date{\today}
\author{Saijun Wu, Thomas Plisson, Roger Brown, William D. Phillips and J. V. Porto}
\affiliation{Joint Quantum Institute, NIST and University of Maryland,
Gaithersburg, Maryland 20899}
\begin{abstract}
We demonstrate a Magneto-Optical Trap (MOT) configuration which employs optical forces due to light scattering between electronically excited states of the atom. With the standard MOT laser beams propagating along the {\it x}- and {\it y}- directions, the laser beams along the {\it z}-direction are at a different wavelength that couples two sets of {\it excited} states. We demonstrate efficient cooling and trapping of cesium atoms in a vapor cell and sub-Doppler cooling on both the red and blue sides of the two-photon resonance. The technique demonstrated in this work may have applications in background-free detection of trapped atoms, and in assisting laser-cooling and trapping of certain atomic species that require cooling lasers at inconvenient wavelengths.\end{abstract}
\pacs{37.10.De, 37.10.Vz, 32.80.Wr}

\maketitle

The development of laser cooling and trapping techniques in the last three decades has greatly enhanced our ability to control atoms, impacting a range of fields from precision atomic measurements and atomic clocks to quantum degenerate gases and quantum information processing. To date, most laser cooling methods use the mechanical effect of single-photon transitions between ground states and electronically excited states. These include Doppler cooling, polarization gradient cooling, and velocity-selective coherent population trapping~\cite{metcalf}. There are, however, a few theoretical and experimental studies involving laser cooling in three-level systems comprising a ground state and two electronically excited states. For example, ref.~\cite{Rooijakkers97} showed an enhancement of radiation pressure by driving a 2-photon transition in a 3-level system. In other work, the effective linewidth for the cooling transition was controlled by dressing the excited state via a coupling to another excited state. This effect can either broaden~\cite{binnewies01a,curtis01a} or narrow~\cite{Malossi05} the effective single-photon cooling transition.

\begin{figure}
\centering
\includegraphics [width=1.6 in,angle=270] {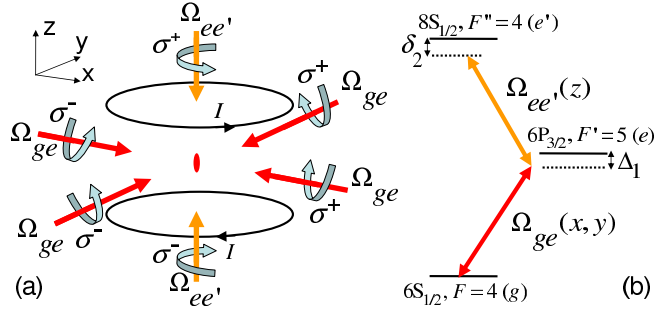}
\caption{(Color online) (a): Schematic of the setup in this work; $\sigma^{\pm}$ are specified with respect to the positive $x$, $y$ and $z$ axes. (b): Simplified level diagram and related transitions.}\label{figSetup}
\end{figure}

Exploiting the Doppler and Zeeman shifts of single-photon optical dipole transitions, the Magneto-Optical Trap (MOT)~\cite{firstMOT} has been the standard tool to cool and trap neutral atoms in 3D. The primary motivation of this work is to use Doppler and Zeeman shifts of multi-photon transitions to both cool and trap atoms. We demonstrate a trap geometry where the cooling and trapping of atoms along one axis of the 3D-trap is due entirely to optical forces from transitions between two {\em electronically excited} states~\cite{foot:exp}. Specifically, with the 852~nm cooling laser beams of a standard cesium($^{133}$Cs) MOT propagating along the $x$- and $y$- directions, we replace the laser beams along the $z$-direction with counter-propagating 795~nm laser beams that only couple the excited states of cesium (6P$_{3/2}$ $F'$=5) to a third set of excited states (8S$_{1/2}$ $F''$=4) (see Fig.~\ref{figSetup}). In this two-color MOT we find efficient cooling along the $z$-direction at both small and large two-photon detunings, while a magneto-optical restoring force was found when the helicities of the 6P-8S beams are opposite to those for the standard MOT.  Remarkably, the two-color MOT can reach sub-Doppler temperatures at both positive and negative two-photon detunings.

The new feature of the two-color MOT sketched in Fig.~\ref{figSetup} is in the cooling and trapping along the $z$- direction. Consider the low intensity regime where the rate of excited atoms leaving from both 6P$_{3/2}$ and 8S$_{1/2}$ states is dominated by spontaneous emission, with negligible contribution from stimulated processes. In this regime, the dominant radiation pressure along $\bf \hat z$ is due to 2-photon scattering, where the first photon is absorbed from the in-plane laser beams and the second is absorbed from the beams along $\bf \hat z$. In particular, we consider $R^{(2)}_{\bf \hat i,\bf \hat j}$, the rate of 2-photon scattering induced by a 6S-6P beam along $\bf \hat i$ and a 6P-8S beam along $\bf \hat j$. Here ${\bf \hat i}\in\{{\bf \hat x},-{\bf \hat x},{\bf \hat y},-{\bf \hat y}\}$ is one of the four directions of the 6S-6P beams, and ${\bf \hat j}\in \{{\bf \hat z},-{\bf \hat z}\}$ is one of the two directions of the 6P-8S beams. The scattering force along $\bf \hat z$ can be written as ${\bf f}^{(2)}_z=\hbar  k_{ee'} \sum\limits_{\bf \hat i,\hat j} R^{(2)}_{\bf \hat i,\hat j}{\bf \hat j}$.  For an atom moving at velocity $\bf v$, we have the 2-photon scattering rate in the low intensity limit:
\begin{equation}
R^{(2)}_{\bf \hat i,\bf \hat j}=\frac{\gamma |\Omega_{ge} \Omega_{ee'}|^2}{16|(\tilde \Delta_1-k_{ge} {\bf \hat i\cdot v})(\tilde \delta_2-k_{ge} {\bf \hat i\cdot v}-k_{ee'} {\bf \hat j\cdot v})|^2}.
\label{2photonrate}
\end{equation}
Here $\Omega_{ge}$ and $\Omega_{ee'}$ are the Rabi frequencies of the laser induced couplings per beam; $k_{ge}$ and $k_{ee'}$ are the wavenumbers of the laser beams; $\tilde \Delta_1=\Delta_1+i \Gamma/2$ and $\tilde \delta_2=\delta_2+i \gamma/2$; $\Delta_1$ and $\delta_2$ are the 1-photon and 2-photon detunings for the 6S$_{1/2}$ $F=4$ to 8S$_{1/2}$ $F'' = 4$ 2-photon excitation, with 6P$_3/2$ $F^\prime=5$ as the intermediate level (Fig.~\ref{figSetup}b); $\Gamma/2\pi=5.2$~MHz and $\gamma/2\pi=1.5$~MHz are the linewidths of the 6P$_{3/2}$ and 8S$_{1/2}$ states respectively.

\begin{figure}
\centering
\includegraphics [width=1.27 in,angle=270] {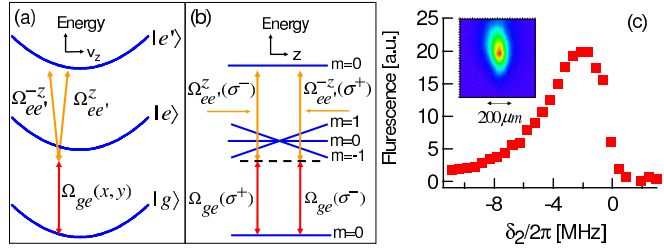}
\caption{(Color online) (a): Schematic illustration of the velocity damping due to the Doppler effect for 2-photon scattering. Here $\Omega_{ee'}^{\pm z}$ represents the 6P-8S beam from the $\pm \bf \hat z$ direction. (b): Schematic illustration of the trapping force along {\it z} due to the Zeeman shift ({\it z} quantization axis) of intermediate resonance in the 2-photon scattering in a linearly changing magnetic field. Only the excitation pathway enhanced by the Zeeman shift is shown. (c): Peak fluorescence of the two-color MOT vs 2-photon detuning $\delta_2$. Here $s_{ge}=1$, $s_{ee'}=15$. Inset gives a fluorescence image of the MOT at $\delta_2/2\pi=-3$~MHz.}\label{fig2ct}
\end{figure}

Taylor-expanding Eq.~(\ref{2photonrate}) around $v_z={\bf \hat z\cdot v}=0$ gives $f^{(2)}_z\approx -\alpha^{(2)} v_z$, with $\alpha^{(2)}>0$ (damping) for negative 2-photon detuning $\delta_2<0$. This 2-photon version of the usual~\cite{metcalf} Doppler cooling mechanism can be summarized with the level diagram in Fig.~\ref{fig2ct}a: the Doppler effect enhances the absorption cross-section for the 6P-8S beam opposing the velocity. One qualitative difference from standard Doppler cooling is that the 2-photon transitions to the 8S states are not closed, so we expect that repumping light will be important to keep the population from pumping into the 6S$_{1/2}$ $F=3$ ground states.

In addition to the velocity-dependent force, a position-dependent restoring force along the $z$-direction is essential for trapping. Figure.~\ref{fig2ct}b illustrates the basic principle of the trapping force. To simplify our discussion, we consider a hypothetical atom with angular momentum $J$=0 ground state, $J'$=1 intermediate states and $J''$=0 excited state. As with the cooling force, the trapping force along $\bf \hat z$ is due to the scattering of the 6P-8S light. The position dependence of this force is due to the spatially dependent Zeeman shift of the {\em intermediate} 6P$_{3/2}$ levels. Taking the quantization axis along $\bf \hat z$, the 6S-6P beams in the $x-y$ plane provide both $\sigma$ and $\pi$ couplings between the ground state and the intermediate states. For a magnetic field along $+\bf \hat z$, ($z>0$: right side of Fig.~\ref{fig2ct}b), the intermediate detuning of the 2-photon excitation is shifted toward resonance for the excitation pathway involving a $\sigma-$ transition to the intermediate state followed by a $\sigma+$ transition to the excited state. As a result, the atoms at $z>0$ preferentially absorb the 6P-8S light propagating toward $-\bf \hat z$, leading to a restoring force in a magnetic quadruple field. Unlike the damping force, this restoring force has the correct sign for both positive and negative $\delta_2$ when $\Delta_1<0$.

The above analysis is corroborated by our experimental observations. In particular, at moderate 6P-8S intensity the 2-photon detuning must be negative to achieve laser cooling along the $z$-direction of the trap. Surprisingly, at high intensities laser cooling and trapping behave differently. As detailed below, we found laser cooling on both the red and blue sides of the 2-photon resonance. We argue that this counter-intuitive effect is due to 3-photon and higher order scattering processes.

Our experiments capture, cool and trap atoms in a cesium vapor cell. The cooling light in the $x-y$ plane comprises the two pairs of counter-propagating 852~nm laser beams (6S-6P beams) with 8~mm $1/e^2$ diameter. (See Fig.~\ref{figSetup}.) The single photon detuning $\Delta_1/2\pi=-12.5$~MHz and the peak intensity of each beam is characterized by $s_{ge}\equiv \frac{2\Omega_{ge}^2}{\Gamma^2}$. The gradient of the magnetic quadruple field was 1.4~mT/cm along $\bf \hat z$.  The beams along $\bf \hat z$ are a pair of 795~nm laser beams (6P-8S beams), and the peak intensity of each 6P-8S beam is characterized by the  parameter $s_{ee'}\equiv \frac{2\Omega_{ee'}^2}{\gamma^2}$~\cite{foot:exp1}. We add two counter-propagating repump beams at 895~nm along $\bf \hat x$, tuned to the  6S$_{1/2}$ $F=3$ to 6P$_{1/2}$ $F^\prime =4$ transition to keep atoms in the $F=4$ ground states.

With the 6P-8S beams at a moderate intensity of 20~mW/cm$^2$ ($s_{ee'}\approx15$, $\Omega_{ee'}/2\pi\approx4$~MHz), and guided by the 2-photon Doppler cooling picture (Fig.~\ref{fig2ct}a), we set the 2-photon detuning $\delta_2$ to small negative values, comparable to the 8S linewidth $\gamma$. We observe trapped atoms in the two-color MOT when the helicities of the 6P-8S beams are set to be opposite to those of the 6S-6P beams in a standard MOT (Fig.~\ref{figSetup}b, Fig.~\ref{fig2ct}b). As with a standard MOT~\cite{shimizu89, monroe90}, we find that our trap tolerates wrong helicity components in the 6P-8S beams with up to $\approx30\%$ in intensity. As expected, the two-color MOT is more sensitive to the repump efficiency than a standard MOT, and the counter-propagating beams need to be intensity-balanced to nullify the repump radiation pressure.

In Fig.~\ref{fig2ct}c we plot the peak fluorescence of the two-color MOT vs $\delta_2$. At the optimal 2-photon detuning of $\delta_2/2\pi\approx-3$~Mhz and with $s_{ge}\approx4$, up to $8\times10^5$ atoms at a density of $5\times 10^{10}$/cm$^3$ are accumulated in the two-color MOT from the pressure P$\approx10^{-5}$~Pascal ($10^{-7}$~Torr) cesium vapor. Due to the weaker trapping and damping along $\bf \hat z$, both the spatial and velocity distributions of the atomic sample are elongated along $\bf \hat z$. The velocity spread of the atoms along $\bf \hat x$ and $\bf \hat z$ is characterized by effective temperatures $T_x\approx70$~$\mu$K and $T_z\approx700$~$\mu$K, both of which are reduced at smaller $s_{ge}$ (see below and Fig.~\ref{figT}).  Typical $1/e^2$ widths of the atomic spatial distribution, fit to a Gaussian, are $w_{x}\approx300$~$\mu$m and $w_z\approx600$~$\mu$m. The number of trapped atoms is an order of magnitude smaller than that of a standard MOT under similar conditions, which is likely due to the reduced capture velocity and effective capture volume for the two-color MOT.


\begin{figure}
\centering
\includegraphics [width=1.65 in,angle=270] {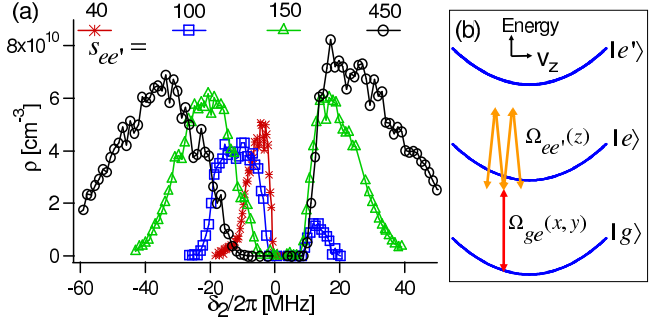}
\caption{(Color online) (a): Peak atom density vs two-photon detuning $\delta_2$ for atoms in the two-color MOT at different 6P-8S beam intensities $s_{ee'}$ and for $s_{ge}=4$. (b): Schematic illustration of the velocity damping due to the Doppler effect for 3-photon scattering. }\label{fig3photon}
\end{figure}

The 2-photon Doppler cooling picture (Fig.~\ref{fig2ct}) fails dramatically at high 6P-8S beam intensities. As $s_{ee'}$ increases, the range of $\delta_2$ for MOT operation broadens and shifts to the red. When $s_{ee'}$ is larger than a threshold value of $s_{\rm th}\approx80$, the two-color MOT also works at positive $\delta_2>\delta_{\rm th}\approx2\pi\times$10~MHz (Fig.~\ref{fig3photon}a). For $s_{ee'}\approx1.8 \times 10^3$ (not shown in Fig.~\ref{fig3photon}), the two-color MOT operates for $\delta_2$ spanning a range more than $2\pi\times$100~MHz ($>>\gamma,\Gamma$) on both the red and blue sides of the two-photon resonance. The maximum number of trapped atoms is similar to that achieved in the low 6P-8S beam intensity regimes, but with up to $50\%$ increase of peak atom densities.

For high 6P-8S beam intensity and moderate 2-photon detuning, both the spatial and velocity distributions of the trapped atomic sample are more isotropic than those at low intensity. As $s_{ee'}$ increases, the ratio $w_z:w_x$ can reach or even go below unity at small positive $\delta_2$. The ratio $T_z:T_x$ decreases and approaches unity as $s_{ee'}$ increases, while a larger $s_{ee'}$ is needed for the same ratio to be reached at a larger $|\delta_2|$. The effective temperature $T_x$, and remarkably, also $T_z$, decrease linearly with $s_{ge}$ until the MOT stops working. For $s_{ge}<1$, $T_z$ is well below the 125~$\mu$K D2 Doppler limit at both large $|\delta_2|$ as well as at small positive 2-photon detunings, as shown in Fig.~\ref{figT}. In addition, at large $|\delta_2|$ the MOT becomes less sensitive to the repump efficiency and intensity balance, as in a standard MOT.

The observation of laser cooling and trapping on the blue side of the 2-photon resonance is intriguing. Equation~(\ref{2photonrate}) indicates that for $\delta_2>0$, the Doppler effect leads to a velocity-dependent force that becomes anti-damping. At low intensity, this precludes operation of the MOT. However, the 2-photon force picture ignores higher order scattering processes, which can be important at high intensities. These include the 3-photon process sketched in Fig.~\ref{fig3photon}b in which a 2-photon absorption is followed by a stimulated emission from 8S to 6P. These multi-photon processes can lead to efficient cooling along $\bf \hat z$ in a manner similar to Doppleron cooling~\cite{Doppleron90}. In the same way as for 2-photon force calculations, the 3-photon scattering force can be written as ${\bf f}^{(3)}_z=2\hbar k_{ee'} \sum\limits_{\bf \hat i,\hat j}  R^{(3)}_{\bf \hat i,\hat j,-\hat j}{\bf \hat j}$, where, for atoms moving at velocity $\bf v$, the 3-photon scattering rate $R^{(3)}_{\bf \hat i,\hat j,-\hat j}$ is

\begin{equation}
R^{(3)}_{\bf \hat i,\bf \hat j,-\bf \hat j}=\frac{|\Omega_{ee'}|^2}{4|\tilde \Delta_1-k_{ge} {\bf \hat i\cdot v}-2 k_{ee'} {\bf \hat j\cdot v}|^2}\frac{\Gamma}{\gamma} R^{(2)}_{\bf \hat i,\bf \hat j}\label{3photonrate},
\end{equation}
with $\tilde \Delta_1$ and $R^{(2)}_{\bf \hat i,\bf \hat j}$ as in Eq.~(\ref{2photonrate}).

As in our treatment of Eq.~(\ref{2photonrate}), we Taylor-expand $f^{(3)}_z$ near $v_z=0$ to find the 3-photon damping coefficient $\alpha^{(3)}$. For $\Delta_1<0$ and $\gamma^2<<\Gamma^2+4\Delta_1^2$, we find $\alpha^{(3)}>0$ for either $\delta_2<0$, or $\delta_2>-\Delta_1/2$. We note that $\alpha^{(3)}$ involves only the 3-photon process and ignores 2-photon processes, light shifts and higher order processes. The 3-photon cooling effect at $\delta_2>0$ can be understood qualitatively from the diagram in Fig.~\ref{fig3photon}b: At large $|\delta_2|$, the Doppler sensitivity along $\bf \hat z$ of the 6P-8S-6P Raman process becomes independent of $\delta_2$, but remains dependent on $\Delta_1$. The fact that $\alpha^{(3)}$ is positive is determined by the negative single-photon detuning $\Delta_1$. In addition, the decreased 8S population at large  $|\delta_2|$ reduces the two-color contribution to unwanted optical pumping into the $F=3$ ground states, which helps explain the decreased sensitivity on repump light.

\begin{figure}
\centering\includegraphics [width=2.75 in,angle=270] {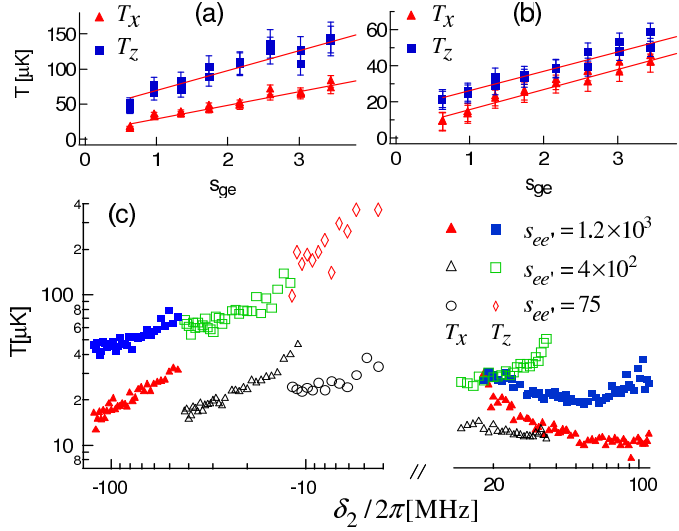}
\caption{(Color online) (a, b): Temperature of the atoms vs. $s_{ge}$ for  $s_{ee'}=1.8\times 10^3$, with $\delta_2/2\pi=-143$~MHz in (a) and $\delta_2/2\pi=117$~MHz in (b). Notice the different temperature scales. (c): Temperature vs $\delta_2$ for atoms in the two-color MOT at various $s_{ee'}$ for $s_{ge}=0.6$.}\label{figT}
\end{figure}

There are at least two possible explanations for the sub-6P$_{3/2}$-Doppler temperatures observed along $\bf \hat z$ over the wide range of 2-photon detunings in Fig.~\ref{figT}. First, as with sub-Doppler cooling in standard optical molasses~\cite{dalibard89}, there is an interplay between spatially dependent light shifts and optical pumping among the 6S Zeeman sublevels, leading directly to sub-Doppler cooling for atoms moving along $\bf \hat z$. This mechanism may be non-intuitive since the 852~nm light, which is the only light field that interacts with the 6S atoms, has no polarization gradient along $\bf \hat z$. However, a $z-$dependent ground state spin polarization can be induced by multi-photon optical pumping processes: the 6P-8S coupling dresses the 6P$_{3/2}$ $F'$=5 Zeeman sublevels, shifting and mixing those sublevels in a $z$-dependent way. An atom excited to a 6P$_{3/2}$ $F'$=5 dressed state is thus spin polarized, and its $z$-dependent polarization is partially retained after the spontaneous decay to the 6S ground states. Combined with a light shift of the ground states due to 2-color processes which is not only $x$-, $y$-dependent but also $z$-dependent, sub-Doppler cooling can occur along $\bf \hat z$. The inseparability of the two-color ground state light shift could also provide a second contribution to the low measured temperature along $\bf \hat z$, by mixing the standard sub-Doppler-cooled motion along $\bf \hat x$, $\bf \hat y$ with the motion along $\bf \hat z$. A quantitative analysis of this ``two-color'' polarization gradient cooling mechanism will appear in a future publication~\cite{longerpaper}.

We have demonstrated a magneto-optical trap where cooling and trapping forces along its $z$-axis are provided entirely by photons associated with transitions between excited states. Up to $8\times10^5$ cesium atoms are trapped in a vapor cell, and the density of the trapped atoms reaches $8 \times 10^{10}$/cm$^3$ at optimal experimental parameters. Sub-Doppler cooling occurs over a wide range of positive and negative 2-photon detunings. Since we observe no density-dependent atom loss, we conclude that two-color-induced collisional loss processes are not particularly large. We believe that the number of atoms in the two-color MOT is lower than that in the standard MOT, because there is a reduced phase-space volume for capture from the room-temperature vapor. We have also observed atom cooling and trapping in a geometry complementary to the setup given by Fig.~\ref{figSetup}a, where the 852~nm beams are along $\bf \hat z$, and the 795~nm beams are along $\bf \hat x$ and $\bf \hat y$, although this geometry traps even fewer atoms.

The two-color cooling and trapping demonstrated here may have practical applications. For instance, a high-numerical-aperture objective can be installed to collect 852~nm fluorescence along $\bf \hat z$ in our setup, a direction along which the scattering of 6S-6P beams from the nearby optics is minimized; the 6P-8S beams at 795~nm wavelength can be easily filtered out. This would enable high-efficiency, near-background-free detection of trapped atoms. This or similar MOT arrangements may also allow completely background-free detection of fluorescence from atomic transitions driven by no laser beam. As another example, replacing regular cooling lasers with excited-state coupling lasers can be technically advantageous for laser cooling of certain atomic species. For example, for atomic hydrogen or anti-hydrogen, the Lyman-$\alpha$ cooling transition needs 121~nm coherent radiation, which is hard to generate and manipulate~\cite{Setija93, Eikema01}. Instead of setting up 3 pairs of Lyman-$\alpha$ beams that couple 1S with 2P for a regular hydrogen MOT, two pairs of the beams may be replaced by laser beams that couple 2P and 3S excited states using the more readily available 656~nm light.

\begin{acknowledgments}
We gratefully acknowledge experimental contributions by Jennifer Sebby-Strabley, and helpful discussions with Vincent Boyer and Bruno Laburthe-Tolra.
\end{acknowledgments}


\begin{thebibliography}{99}
\bibitem{foot:exp}
This is in contrast to ref.~\cite{Malossi05} where the optical force due to excited state couplings is negnigible.
\bibitem{foot:exp1}
$\Omega_{ge}$ and $\Omega_{ee'}$ are speficied for $m_F=4$ -- $m_{F'}=5$ and $m_{F'}=5$ -- $m_{F''}=4$ transitions respectively. We obtain $s_{ge}$ and $s_{ee'}$ by comparing the intensities of the laser beams with effective saturation intensities of the relevant transitions, with $I_0=1.10$~mW/cm$^2$ for $s_{ge}$ and $I_0=1.32$~mW/cm$^2$ for $s_{ee'}$.
\end{thebibliography}
\end{document}